\documentclass[10pt,twocolumn,showpacs,prb,superscriptaddress,aps,floatfix]{revtex4-2}

\usepackage{graphicx}
\usepackage{dcolumn}
\usepackage{bm}
\usepackage{nicefrac}

\usepackage{comment}
\usepackage{gensymb}  
\usepackage{amsmath}
\usepackage{xcolor}
\usepackage{amssymb}
\usepackage{soul}
\usepackage{float}

\newcommand{\cinam}{CNRS/Aix-Marseille Universit\'e, Centre Interdisciplinaire de Nanoscience de Marseille UMR 7325 Campus de Luminy, 13288 Marseille Cedex 9, France}
\newcommand{\qub}{School of Mathematics and Physics, Queen's University Belfast, Belfast BT7 1NN, Northern Ireland, UK}
\newcommand{\etsf}{European Theoretical Spectroscopy Facilities (ETSF)}

\newcommand{\china}{School of Physical Science and Technology and Jiangsu Key Laboratory of Frontier Material Physics and Devices, Soochow University, Suzhou 215006, People’s Republic of China}

\graphicspath{{./Figures/}}

\begin{document}
\newcommand{\baa}{{\bf a}}
\newcommand{\bb}{{\bf b}}
\newcommand{\uu}{{\bf u}}
\newcommand{\qq}{{\bf q}}
\newcommand{\hh}{{\bf h}}
\def\jj{{\bf J}}
\newcommand{\rr}{{\bf r}}
\newcommand{\pp}{{\bf p}}
\newcommand{\PP}{{\bf P}}
\newcommand{\kk}{{\bf k}}
\newcommand{\HH}{{\bf H}}
\newcommand{\GG}{{\bf G}}
\newcommand{\SiS}{{\bf \Sigma}}
\newcommand{\VV}{{\bf V}}
\newcommand{\UU}{{\bf U}}
\newcommand{\w}{\omega}
\newcommand{\tf}{\textbf}
\newcommand{\bo}{\mathbf}
\newcommand{\br}{{\bf r}}
\newcommand{\be}{\begin{equation}}
\newcommand{\ee}{\end{equation}}
\newcommand{\ben}{\begin{equation*}}
\newcommand{\een}{\end{equation*}}
\newcommand{\bea}{\begin{eqnarray}}
\newcommand{\eea}{\end{eqnarray}}
\newcommand{\bean}{\begin{eqnarray*}}
\newcommand{\eean}{\end{eqnarray*}}
\newcommand{\nup}{n_{\uparrow}}
\newcommand{\ndown}{n_{\downarrow}}
\newcommand{\Id}[1] {\int \! \! {\rm d}^3 #1}
\renewcommand{\v}[1]{{\bf #1}}
\renewcommand{\[}{\left[}
\renewcommand{\]}{\right]}
\renewcommand{\(}{\left(}
\renewcommand{\)}{\right)}
\def\efield{\boldsymbol{\cal E}} 
\def\ket#1{\vert#1\rangle}
\def\bra#1{\langle#1\vert}
\def\susc#1{\chi^{(#1)}}
\def\ket#1{\vert#1\rangle}
\def\bra#1{\langle#1\vert}
\def\ai{\emph{ab-initio}\ }

\title{Shift current in 2D Janus Transition-Metal Dichalcogenides: the role of excitons
}

\author{Yuncheng Mao}
\affiliation{\cinam}
\author{Ju Zhou}
\affiliation{\qub}
\affiliation{\china}
\author{Myrta Gr\"uning}
\affiliation{\qub}
\affiliation{\etsf}
\author{Claudio Attaccalite}
\affiliation{\cinam}
\affiliation{\etsf}

\begin{abstract}
We investigate the shift current in two-dimensional (2D) Janus transition-metal dichalcogenides (TMDs). The shift current is evaluated using a real-time approach, where the coupling with an external field is described in terms of a dynamical Berry phase. This methodology incorporates electron-hole interactions and quasiparticle band structure renormalization through an effective Hamiltonian derived from many-body perturbation theory. We find that the shift current is strongly enhanced in correspondence with C excitons. An analysis in terms of the electron-hole pairs reveals that electron and hole are localized on different atoms, and thus, following an optical excitation, the center of the electron charge is displaced, giving rise to a significant photocurrent. Janus TMDs, with their intrinsic out-of-plane asymmetry and tunable electronic properties, are particularly appealing for next-generation optoelectronic and energy-harvesting technologies. These results highlight the role of excitons in the shift-current response of Janus TMDs and demonstrate their potential as promising building blocks for future photovoltaic devices.

\end{abstract}

\maketitle

\section{Introduction}
	The shift current (SC) is a direct current generated in non-centrosymmetric materials under continuous wave illumination, a second-order bulk photovoltaic effect arising from the nonlinear light-matter interaction. This phenomenon presents a promising mechanism for next-generation solar energy conversion, as it generates a photocurrent without the need for p-n junctions. However, the practical application of SC is hindered by typically low short-circuit currents, which are intrinsically linked to solar cell power conversion efficiency. Overcoming this limitation requires a deeper understanding of the underlying physics and the identification of materials with high SC conductivity.

In a recent work, Cook et al.\cite{cook2017design} put forward designing principles for optimizing the SC. They found that low dimensional systems can display a very strong SC due to van Hove singularities leading to a strong density of states. This theoretical prediction is supported by experiments, such as the observation of a strong SC in WS$_2$ nanotubes.\cite{zhang2019enhanced} However, a significant discrepancy persists between these experimental measurements and standard theoretical predictions.\cite{zhang2019enhanced} This disagreement has been attributed to extrinsic factors like tube-tube interactions or incorrect measurement of the tube cross-sections.\cite{PhysRevB.108.165418,kim2022giant} Crucially, most theoretical models overlook a key intrinsic factor: strong electron-hole interactions.

In low-dimensional materials, reduced dielectric screening and quantum confinement lead to the formation of bound excitons, which are known to dramatically reshape the linear and nonlinear optical response.\cite{chan2021giant,hipolito2016nonlinear,esteve2025excitons,morimoto,fei2020shift}
 Excitonic effects could, in principle, strongly modify---and potentially enhance---the SC, yet they are absent from most state-of-the-art calculations.\cite{PhysRevB.108.165418} The lack of these essential many-body effects is likely a primary reason for the theory-experiment gap, underscoring the critical need for a computational approach that incorporates excitonic interactions to accurately predict the SC.

In this work, we address this challenge by presenting a first-principles real-time approach to calculate the SC conductivity. Our method describes the light-matter interaction via a dynamical Berry phase and is built upon a Hamiltonian derived from many-body perturbation theory. This framework allows us to include quasiparticle corrections to the electronic band structure\cite{aryasetiawan1998gw} and to incorporate both local-field effects and electron-hole interactions in the response functions.\cite{strinati} We have previously successfully applied this methodology\cite{attaccalite2013nonlinear,gruning2014second} to other nonlinear optical phenomena, including second harmonic generation and sum-frequency generation (see e.g. \cite{grillo2024tunable,attaccalite2018two,pionteck2025sum}).  Here, we extend it to compute nonlinear conductivities directly from the real-time polarization current.

We apply this approach to investigate the SC in two Janus transition metal dichalcogenides (TMDs): MoSSe and WSSe.\cite{lu2017janus,acsnano.7b03186, acsnano.9b10196} 
These two Janus TMDs have been recently synthesized in the form of monolayers and nanotubes.\cite{lu2017janus,acsnano.7b03186, acsnano.9b10196} 
 Janus monolayers are a novel class of 2D materials where the broken out-of-plane symmetry creates a large built-in electric field,\cite{acs.nanolett.2c04724}  leading to unique properties like a strong piezoelectric response, long exciton lifetimes.\cite{acs.nanolett.2c04724} Their internal electric field renders Janus materials ideal for photocatalysis~\cite{PhysRevB.98.165424}, and their stacking-dependent dipole moment can be used to tune these intrinsic properties.\cite{acs.nanolett.2c04724}
 This $z$-symmetry breaking also makes them ideal candidates for the SC, as it allows for new tensor components, including an out-of-plane response.\cite{strasser2022nonlinear,cpl_40_8_087201}  Furthermore, swapping S with Se atoms switches the chirality of these structure and this property can be used to design heterostructures that enhance the photocurrent.\cite{strasser2022nonlinear}
In this work we focus specifically on the dominant in-plane tensor component, $\sigma^{(2)}_{yyy}$. 
 Regarding the in-plane symmetries (xy-plane), though there are non zero elements of the second-order conductivity tensor, no net SC results for unpolarized light---such as sunlight. It has been shown, that a significant SC can be obtained by breaking the C$_{3v}$ symmetry either by strain, interaction with a substrate\cite{hung2023nonlinear,cpl_40_8_087201}, or by forming nanotubes.\cite{yang2025janus}

The paper is organized as follows: in Section~\ref{sec1} we present our theoretical approach and give all computation details of our calculations; in Sec.~\ref{sec2} results for MoSSe and WSSe are presented with an analysis of the excitonic state responsible for the SC enhancement; finally in Sec.~\ref{sec3} we draw the conclusion and discuss further possible applications.

Our results reveal that excitonic effects cause a profound redistribution of the SC intensity: they strongly suppress the response at the A/B exciton resonance while dramatically enhancing it at the C-exciton resonance. By analyzing the real-space charge distribution of these excitons, we explain this stark contrast as due to the different spatial separation between electron and hole, leading to a substantial shift in the charge center upon photoexcitation—the fundamental driver of the SC.\\ 
With this work we not only resolve key aspects of the SC mechanism in 2D materials but also establish a robust computational framework for its accurate prediction, guiding the search for high-performance bulk photovoltaic materials.

\section{Theoretical methods}
\label{sec1}
We present the definition of the SC response (Sec.~\ref{sec:defSC}) and the real-time approach used to calculate it, as well as the procedures to extract nonlinear conductivity coefficients from the real-time current  (Sec.~\ref{sec:eom}). Then, we detail all the computational parameters used in the simulations  (Sec.~\ref{sec:compdet}).
As these are the first calculations of the SC with the Yambo code\cite{yambo}, in Appendix~\ref{app2} we detail the numerical implementation of the current operator. For validation, we compare with existing results on the GeS monolayer\cite{chan2021giant,PhysRevB.97.245143,esteve2025excitons,PhysRevB.94.155428,PhysRevMaterials.2.034603} (see Supplemental  Material).

\subsection{The shift current}\label{sec:defSC}
When a material is irradiated by light, a current is produced. In the perturbative regime, this current can be expressed as an order expansion of the total field strength:
\begin{equation}
J(\omega)=\sigma^{(1)} (\omega) E(\omega) + \sum_{\boldsymbol{\omega}} \sigma^{(2)}(\omega;\omega_1,\omega_2) E(\omega_1)E(\omega_2) + O(E^3),     
\end{equation}
where the summation on $\boldsymbol{\omega}$ indicates the sum over all distinct $\omega_1,\omega_2$ pairs.  
With $\omega_1, \omega_2 \neq 0$, the lowest order that produces DC current is the second order when $\omega = \omega_1 + \omega_2 = 0$.~\cite{PhysRevB.61.5337} In case of a monochromatic field,
\begin{equation}
J^{(2)}(\omega=0)= 2\sigma^{(2)}(0;\omega_1,-\omega_1) E(\omega_1)E(-\omega_1).
\end{equation}

\subsection{Equation of motion}\label{sec:eom}
We simulate a system under the influence of an external, time-dependent electric field by solving an effective Schr\"odinger equation:
\be
i\hbar  \frac{d}{dt}| v_{\kk,m} \rangle = \left(\hat H_\kk + i {\bf \cal{E}}\cdot \partial_\kk  \right)| v_{\kk,m} (t) \rangle, \label{eom}
\ee
where $\hat H_\kk$ is the effective Hamiltonian, $| v_{\kk,m} (t) \rangle$ are the time-dependent valence states and the dipole operator has been replaced with $\rr = i \partial_\kk$ and the k-derivative has been performed using a covariant approach.\cite{souza_prb,attaccalite2013nonlinear} In the following, we will omit the time-dependence of $| v_{\kk,m} (t) \rangle$ in order to simplify the notation. From the real-time dynamics of the valence states we calculate the current as:
\be
J=\frac{f}{(2 \pi)^3} \sum_{n=1}^M  \langle v_{\kk n}|\hat H_\kk \partial_\kk | v_{\kk n}\rangle + c.c.
\label{eq:current}
\ee
The derivative in $\kk$ space is evaluated by means of finite-difference covariant formulation (see Souza et al. in Ref.~\onlinecite{souza_prb}):
\bea
\tilde \partial_\kk  | v_{\kk n}\rangle &\simeq& \frac{4}{3} \frac{ | \tilde v_{\kk +\Delta \kk, n}\rangle -  | \tilde v_{\kk -\Delta \kk, n} \rangle } {2\Delta \kk} \nonumber \\
&-& \frac{1}{3} \frac{  | \tilde v_{\kk +2 \Delta  \kk, n}\rangle - | \tilde v_{\kk -2 \Delta \kk, n}\rangle }{4\Delta \kk} + O(\Delta \kk)^4,
\label{eq:current2}
\eea
where the tilde indicates that we used the covariant derivative in order to deal with the arbitrary phase factor associated with wave functions at different $\kk$-points.\cite{souza_prb,attaccalite2013nonlinear} In Appendix~\ref{app2}, we describe the numerical implementation of Eqs.~\eqref{eq:current} and \eqref{eq:current2} in the Yambo code~\cite{yambo}.\\
In Eq~\eqref{eom}, the effective Hamiltonian is chosen depending on the level of accuracy we aim at, from independent-particle approximation (IPA) to the time-dependent adiabatic GW (TD-aGW).\cite{attaccalite2011}
The latter approximation is equivalent to the GW+BSE method in the linear response limit.\cite{attaccalite2011,strinati}\\
To obtain the second order conductivity corresponding to the SC,  we perturb the system with a sinusoidal electric field $E(t)=E_0 \theta(t-t_0)sin(\omega_0 t)$, where $E_0,\omega_0$ are the strength and frequency of the applied electric field, respectively, and $\theta$ is the Heaviside step function. Further, we add a term to the Hamiltonian to dephase the eigenmodes that are excited when the field switch-on at $t_0$.\cite{attaccalite2013nonlinear} After transient effects have decayed, the final current can be written as a Fourier series expansion: $J(t) = \sum_{j=-N}^N c_j e^{\mathrm{i}j \omega_0 t}$. The coefficients $c_j$ can be obtained by a discrete Fourier transform by sampling $J(t)$ on a period $2\pi/\omega_0$. The components of the $n$-order conductivity tensor are directly related to the $c_j$ coefficients through powers of the applied electric field. This procedure is the analogue of that used to extract the nonlinear susceptibilities from the polarization density that has been detailed in Refs.~\onlinecite{attaccalite2013nonlinear,pionteck2025sum}. In this case, we extract the conductivity corresponding to the SC from $c_0 = 2\sigma^{(2)}(0;\omega_0,-\omega_0) E_0^2 + O(E_0^4)$.

\begin{figure}[!ht]
   \centering
    \includegraphics[width=0.45\textwidth]{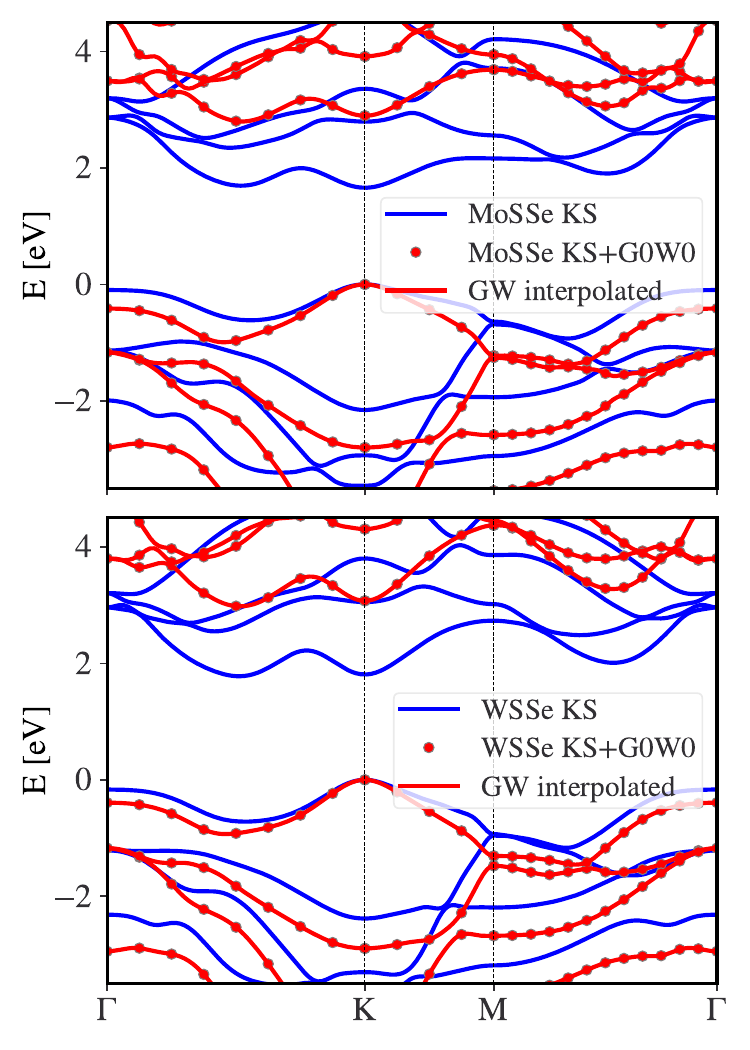}
	      \caption{The electronic band structures of (a) MoSSe and (b) WSSe monolayers, at the Kohn-Sham level (blue solid lines), including the quasiparticle corrections from $\mathrm{G_0W_0}$ calculations (red dots) and interpolated bands (red lines).. See Supplemental Material for the coordinates of the high-symmetry points used.}
 \end{figure}

\subsection{Computational details}\label{sec:compdet}
The ground states of all the systems studied in this manuscript were calculated using Density Functional Theory (DFT) with the PBE functionals and the DOJO stringent pseudo-potential v4.1 to replace the core electrons.\cite{PseudoDojo} We include also a van der Waals correction to the DFT functional.\cite{grimme2006semiempirical} All DFT calculations were performed with the Quantum ESPRESSO code.\cite{pwscf} The linear and nonlinear optical response and quasiparticle corrections were calculated using the Yambo code~\cite{yambo}. For the quasiparticle correction, the G$_0$W$_0$ approximation\cite{aryasetiawan1998gw} was applied, where the screened interaction was treated with the Godby-Needs plasmon-pole model (PPM).\cite{stankovski2011g}

\begin{center}
\begin{table}[h!]
\begin{tabular}{|c|c|c|c|c|c|c|}
	       \hline
        System & $\mathbf{k}$-points & $N_\mathrm{b}$ & $\epsilon_\text{cut}$(Ha) & $\epsilon_\text{bands}$   & $L_z$ (a.u.) & $d_\text{eff}$ (\AA) \\ \hline
	MoSSe      &  $24\times24$ & 19-26 & 5& 400  & 36.30 &6.69   \\ \hline
	WSSe      &  $24\times24$  & 19-26 & 5 & 400  & 36.30 & 6.49   \\ \hline
\end{tabular}
	\caption{The parameters used in the nonlinear response calculations for  MoSSe and WSSe monolayers are: $\kk$-point sampling; the range of bands for the response functions; the cut-off,$\epsilon_\text{cut}$, and the number of bands, $\epsilon_\text{bands}$ used to converge the dielectric function $\epsilon_{\bf G,\bf {G'}}$, and expansion of the Green's function in the GW calculations; the height of the supercell, $L_z$ and the effective layer thickness, $d_\text{eff}$.  As effective layer thickness for MoSSe and WSSe, we used the interlayer distance of the bulk MoS$_2$ and WS$_2$ respectively. \label{table1}}
\end{table}
\end{center}
We calculated the optical absorption at the independent particle approximation (IPA) on top of Kohn-Sham band structure, adding G$_0$W$_0$ corrections (IPA-GW) and including the local-field effects and electron-hole interaction by solving the Bethe-Salpeter equation in the Tamm-Dancoff\cite{strinati} (GW+BSE). 
While the IPA and IPA-GW are solved within the linear response framework, the GW+BSE is formulated as an eigenvalue problem for the two-particle Hamiltonian:
\bea
\label{eq:H_BSE}
H_{\substack{vc\boldsymbol{k}\\v'c'\boldsymbol{k}'}}=\left(E_{c\boldsymbol{k}}-E_{v\boldsymbol{k}}\right)\delta_{vv'}\delta_{cc'}\delta_{\boldsymbol{k}\boldsymbol{k}'}+ \nonumber \\
+\left(f_{c\boldsymbol{k}}-f_{v\boldsymbol{k}} \right)\left(2\overline{V}_{\substack{vc\kk\\v'c'\kk'}} - W_{\substack{vc\boldsymbol{k}\\v'c'\boldsymbol{k}'}}\right),
\eea
where $E_{n\boldsymbol{k}}$ are quasiparticle energies and $f_{n\boldsymbol{k}}$ the electronic occupations and $n$ run over the valence ($v$) and conduction bands ($c$). $\overline{V}$ is the Coulomb potential derived from the variation of the Hartree term and $W$ the screened electron-hole interaction derived from the screened exchange.\cite{strinati}
The eigenvectors $A^{\lambda}_{cv\boldsymbol{k}}$ and eigenvalues $E_{\lambda} $ of the Hamiltonian from Eq.~\ref{eq:H_BSE} are respectively the excitonic wave-functions and energies. These quantities are used to build up the macroscopic dielectric function $\epsilon_{M}$ which writes:
\begin{equation}
\epsilon_M(\omega)= 1 - 4\pi\sum_\lambda \frac{ | T_\lambda|^2}{\omega - E_\lambda + i\eta},
\label{eq:epsilon}
\end{equation}
where $T_\lambda =  \sum_{cv\kk} d_{cv\kk} A^\lambda_{cv\kk}$ are the excitonic dipoles, $d_{cv\kk}$ are the dipole matrix elements between the Kohn-Sham states with $d_{cv\kk} = \langle v \kk | \hat r | c\kk \rangle$, and $i\eta$ with $\eta$ chosen to be 0.1~eV is a small broadening factor to simulate the experimental spectra. In order to consider isolated monolayers, a cutoff was applied to the Coulomb potential to avoid interaction between periodic replicas along the $z$-direction (Rozzi et al. \onlinecite{rozzi2006exact}). Parameters used in the $G_0W_0$ and BSE (Tab.~\ref{table1}) are sufficient to have band gaps converged within 0.05~eV and exciton energies within 0.001~eV due to the error cancellation between band gap and excitonic binding energy (see also Sec.~III in  Supplemental Material).  The $G_0W_0$ corrections used in the BSE and the real-time propagation are calculated directly on a regular grid using such convergence criteria. Exclusively for plotting, the $G_0W_0$ band structure is obtained by interpolation using smooth Fourier transform.\cite{SKW}\\

The nonlinear response is evaluated from the solution of the equation of motion as in Sec.~\ref{sec:eom} at the corresponding levels of theory as for the optical absorption\cite{attaccalite2011}: TD-IP, TD-IP@GW, and TD-aGW. The equations of motion have been solved using the Crank-Nicolson solver\cite{attaccalite2013nonlinear}, with an integration time of 83~fs, a time-step of 0.01~fs and a dephasing time corresponding to a 0.1~eV spectral broadening. The nonlinear response was extracted from the real-time current using the YamboPy code\cite{YamboPy}. All other relevant computational details are presented in Table~\ref{table1}. 
The spin-orbit coupling was not included in the calculations, as it has been shown to have little effect on the nonlinear response of Janus TMDs~\cite{strasser2022nonlinear}. 

\section{Results}
\subsection{Atomic and electronic structure}
\label{sec2}
For both MoSSe and WSSe we optimize atomic positions and unit cell at zero pressure by minimizing the enthalpy as implemented in the Quantum Espresso code.\cite{pwscf} We obtained lattice parameters of  6.089~a.u. and 6.090~a.u. respectively for MoSSe and WSSe (lattice vectors and atomic coordinates are specified in the Supplemental Material). Starting from the optimized atomic structure we calculate the electronic bands and then the linear and nonlinear response functions.
The electronic band structures are evaluated at the Kohn-Sham (KS) and G$_0$W$_0$ levels for both MoSSe and WSSe, results are shown in Fig.~\ref{fig:TMD_bands}.

\begin{figure}[t]
   \centering
    \includegraphics[width=0.5\textwidth]{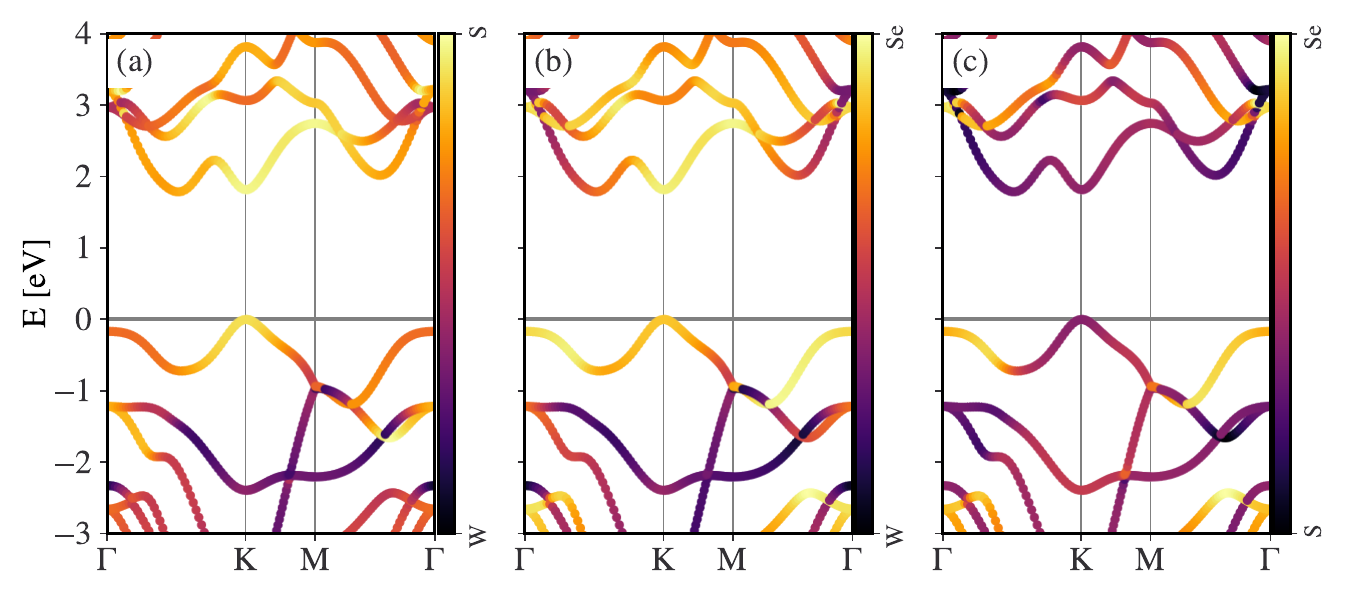}
	      \caption{Projected band structure of WSSe on atomic orbitals. Panel (a): projection weight onto S and W; panel (b): projection weight onto Se and W; panel (c): projection weight onto S and Se.}\label{fig:WSSeproj}
      \end{figure}

For MoSSe, we found a gap of 1.66~eV at the PBE level and 2.80~eV in the G$_0$W$_0$ approximation. For WSSe, the KS gap is 1.78~eV while the G$_0$W$_0$ corrections increase it to 2.98 eV. These results are in agreement with previous calculations.\cite{PhysRevB.104.125306,li2019excited,li2023direct} Interestingly, there is a strong $k$-dependence of the G$_0$W$_0$ corrections for the valence bands.
Note that the band gaps in these materials are highly sensitive to calculation details and strain, as shown in Ref.~\onlinecite{hou2020room}, and small differences between computational approaches can cause a transition from direct to an indirect semiconductor in the monolayers. The direct/indirect transition affects light-emission properties, but do not impact significantly light absorption or the SC, which only involves direct electron-hole transitions.
Figure~\ref{fig:WSSeproj} shows the band structure of WSSe including information on the atomic orbital nature of the electronic wave functions. Of particular interest are the bands closest to the Fermi energy, as these contribute to the main features in the shift current response. The states in the top valence and bottom conduction bands are mostly localized on the chalcogen atoms (panel a and b), with the exception for states located close to the minima along the M$\Gamma$ direction which are mostly localized on the W atom. On the top valence band, states at and around $\Gamma$ are mostly localized on the Se atoms while those around K are mostly localized on the S atoms. The bottom conduction band has mostly an S atom character except around the $\Gamma$ point. From the corresponding analysis of the band structure in terms of the  atomic orbital nature of the electronic wave functions for MoSSe (see Supplemental Material), we conclude that the localization of the closest states to the Fermi energy is substantially the same as in WSSe.

\subsection{Linear and nonlinear response in MoSSe and WSSe}
\begin{figure}[!ht]
   \centering
    \includegraphics[width=0.45\textwidth]{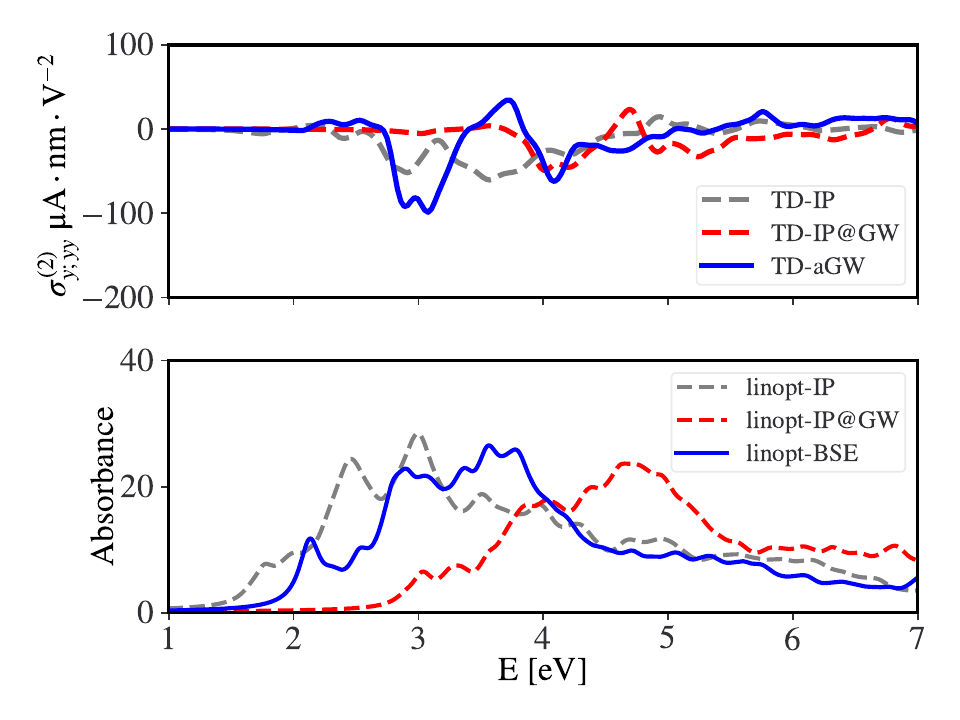}
	      \caption{Shift current along the $yyy$ direction for MoSSe (upper panel) compared with the linear absorption (lower panel), at the TD-IP (gray dashed lines), TD-IP@GW (red dashed lines) and TD-aGW(GW+BSE) (blue solid lines) levels.\label{fig:MoSSe_NLBSE}}
 \end{figure}

For the two systems, we first present the linear optical absorption, which later serves as a guide for interpreting the SC conductivity. In the bottom panels of Figs.~\ref{fig:MoSSe_NLBSE} and~\ref{fig:WSSe_NLBSE} we report the optical absorbance at different levels of approximation: independent particle (IPA), independent particle on top of the GW band structure (IPA-GW), and the GW+BSE approximation.\\

\begin{figure}[!ht]
   \centering
    \includegraphics[width=0.5\textwidth]{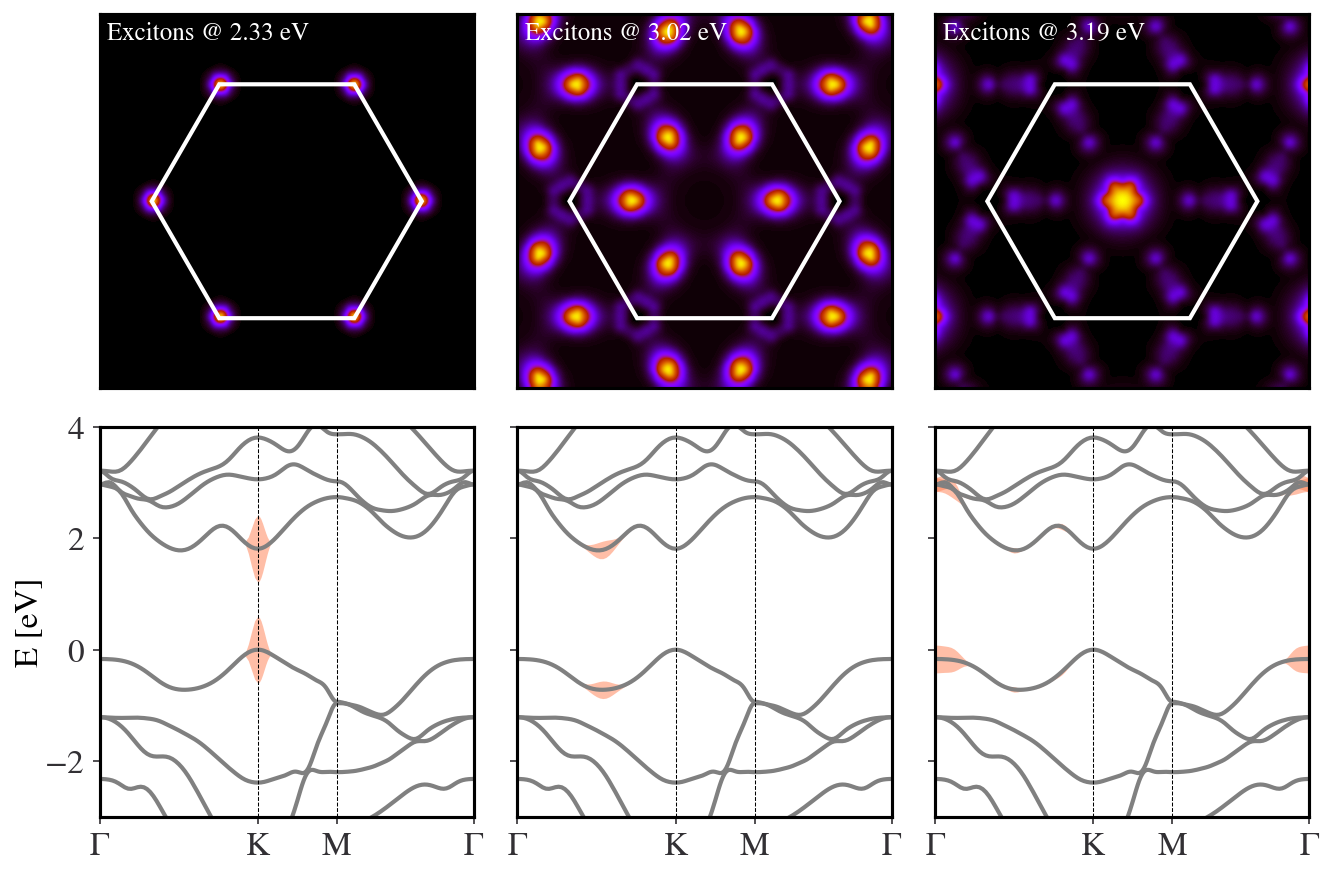}
	      \caption{Exciton analysis in WSSe. Top left (right) panel: the projection onto the BZ of the A/B, C$_1$ and C$_2$ excitons. Bottom left (right) panel: the projection along the WSSe band structure of the A/B, C$_1$ and C$_2$ excitons.\label{exc_analysis}}
 \end{figure}

\begin{figure}[!ht]
   \centering
    \includegraphics[width=0.5\textwidth]{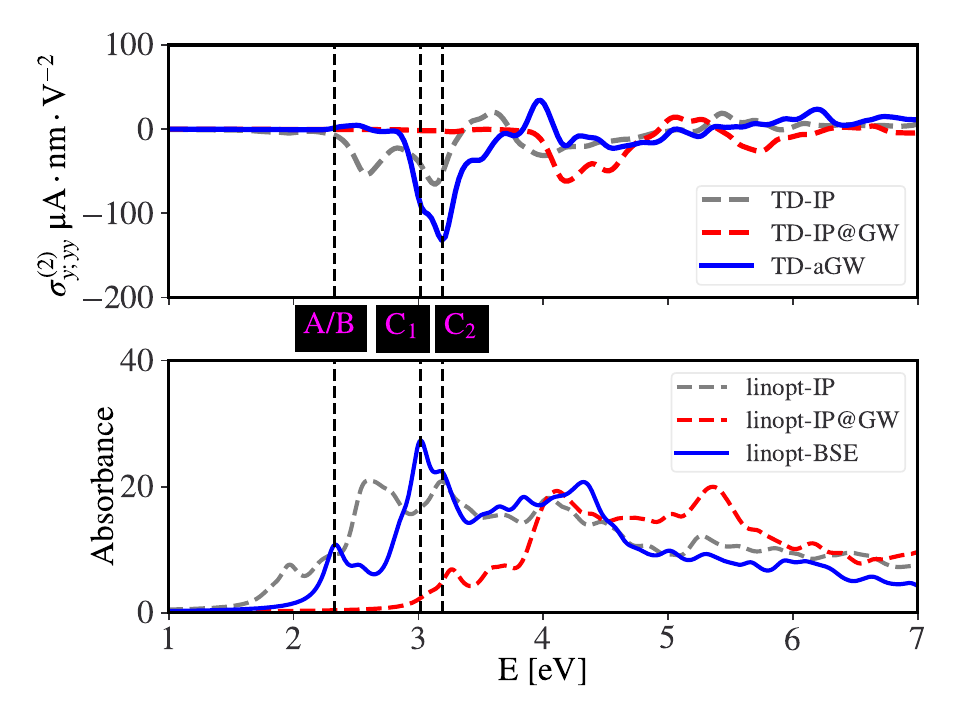}
	      \caption{Shift current along the $yyy$ direction in WSSe (upper panel) compared with the linear absorption (lower panel), at the TD-IP (gray dashed lines), TD-IP@GW (red dashed lines) and TD-aGW (GW+BSE) (blue solid lines) levels.\label{fig:WSSe_NLBSE}. The positions of A/B, C1 and C2 excitons analyzed in Fig.~\ref{exc_analysis} are labeled with magenta letters in black boxes.}
 \end{figure}

In Figs.~\ref{fig:MoSSe_NLBSE} and~\ref{fig:WSSe_NLBSE} we can distinguish different excitonic peaks for both MoSSe and WSSe: in particular the A/B peak around 2~eV and the C peaks around the 3~eV. Notice that as we did not include spin-orbit coupling, the ground state exciton (A) and the spin-orbit split exciton (B) are degenerate. The strongest features in both spectra are the C-excitons, which we found to correspond to the strongest enhancement of the SC. For this reason we analyze them in detail.\\
We consider the two most intense excitations, C$_1$ at 3.02~eV and C$_2$ at 3.19~eV for WSSe, together with the A/B exciton at 2.33~eV, indicated with vertical lines in Fig.~\ref{fig:WSSe_NLBSE}. We then projected their exciton weights on $k$-points in the full Brillouin zone $W^\lambda_\kk=\sum_{cv}|A^\lambda_{cv\kk}|^2$, top panel of  Fig.~\ref{exc_analysis}, and on the valence/conduction bands along the band structure of WSSe, $W^\lambda_{c/v,\kk}=\sum_{v/c} |A^\lambda_{cv\kk}|^2$, bottom panel of Fig.~\ref{exc_analysis}. 
The A/B exciton originated from electron-hole transitions around K,
the $C_1$ exciton originated from electron-hole transitions along  K-$\Gamma$, the $C_2$ exciton from electron-hole transitions around $\Gamma$. For all three excitons, the holes are in the top valence band and the electrons in the bottom conduction band.  

By comparing the distribution of the $W^\lambda_{c/v,\kk}$ with the atomic orbital projected band structures (see Supplemental Material),
we can deduce the spatial localization of the electron-hole pairs contributing to each exciton. For the A/B exciton, both electrons and holes are mostly localized on the chalcogen atoms. Being the composition of the bottom conduction and top valence around K very similar, we deduce the electron and hole are localized around the same atom. For the $C_1$ exciton, while the electrons are mostly localized around the S atoms, the holes are partially localized on the Se atoms as well. For the $C_2$ exciton, holes are more localized on the tungsten and electrons more on the S atoms.

A similar analysis holds for the C-excitons around 3~eV of MoSSe (see Supplemental Material). Note that, for MoSSe, the most intense features are at higher energies than the C excitons. Nevertheless, as these features are out of visible range and so not relevant for photovoltaic applications, we do not provide an analysis of these excited states. \\

Next, we consider the SC. Symmetry analysis shows that SC in these Janus dichalcogenides has four independent tensor elements.\cite{strasser2022nonlinear} Three of these tensor elements involve out-of-plane directions $\sigma^{(2)}_{z;xx}$,$\sigma^{(2)}_{z;yy}$, $\sigma^{(2)}_{z;zz}$, while the fourth involves only in-plane components $\sigma^{(2)}_{y;yy}$. In this work, we focus specifically on the fourth component of the SC tensor for three reasons. First, it is the largest component~\cite{strasser2022nonlinear}. Second, the in-plane response is less sensitive to substrate interaction and therefore easier to measure. Third, our method, which is based on the dynamical Berry phase, is limited to periodic systems, and we are still working to include non-periodic directions in our real-time approach.

Figures~\ref{fig:WSSe_NLBSE} and~\ref{fig:MoSSe_NLBSE} show the SC of MoSSe and WSSe, respectively, at the  three levels of approximation: TD-IP, TD-IP@GW, and TD-aGW. These correspond to independent-particle, independent-particle on top of $G_0W_0$ band structure, and GW+BSE, respectively.\cite{attaccalite2011}
In the nonlinear response function, the gap widening induced by $G_0W_0$ corrections substantially modifies the spectral intensity by suppressing the SC conductivity. Including excitonic effects redshifts the full spectrum and restores its intensity, albeit with a significant redistribution with respect to the TD-IP. In both systems, the SC conductivity is strongly suppressed in correspondence of the A/B exciton, while it is strongly enhanced in correspondence of the C-excitons. A similar intensity redistribution has been observed in other 2D dichalcogenides in second-harmonic generation\cite{acs.nanolett.4c03434} and sum-frequency generation.\cite{pionteck2025sum}  The relation between second harmonic generation (SHG) and SC is also discussed in the work of Morimoto and Nagaosa in Ref.~\onlinecite{morimoto}. 

To explain the stark difference of the SC at resonance with the A/B and C excitons, we refer to the analysis in terms of electron-hole pairs. While for the A/B exciton the holes and electrons are localized around the same atom, for the C excitons the electron and holes are localized on different sites. We thus argue that following a photo-excitation at resonance with the C excitons there is a shift of the center of electronic charge, which is the physical origin of the SC.    

We now highlight the key differences between the two systems, MoSSe and WSSe. Two important differences stand out as we move from one to the other: the first is the different band gaps, reflected in the shift of the SC spectra. The WSSe spectra starts at slightly higher energies than that of MoSSe. Second, in WSSe, C-excitons are more intense and better separated from higher excitons than in MoSSe. These differences will be important for selecting the best material having a response in the desired energy range in experimental devices. 
Finally, we calculate the total short-circuit current for the two system studied here (see Sec.~VI in Supplemental Material), and we found a maximum photo-current of 1.2~nA at the C resonance of the WSSe monolayer. This value is much larger than the  maximum reported value for bulk BaTiO$_3$\cite{PhysRevLett.109.116601} and  is 20\% larger than the one reported for monolayer WS$_2$.\cite{PhysRevB.108.165418} 

\section{Conclusions}
\label{sec3}
We studied the SC response of two-dimensional (2D) Janus dichalcogenides, MoSSe and WSSe, using a real-time approach that includes quasi-particle corrections in the electronic band structures and takes into account local field effects and electron-hole interaction in the response functions. We found that, in both systems, excitonic effects strongly enhance the SC conductivity at resonance with C-excitons, while suppresses it at resonance with the A/B exciton. As the SC originates from the shift of the center of electron charge following photo-excitations, we rationalize this result by noting that in the A/B exciton, the electrons and holes are localized around the same atomic site---thus no  shift of the center of electron charge---while in the C excitons, the real-space distribution of electrons and holes differs significantly, so following a photoexcitation resonant with the C excitons, there is a shift of the center of electron charge and thus a photocurrent.        
To harness this significant response in devices, the $\mathrm{C_{3V}}$ symmetry must be broken. This can be achieved by combining the materials into heterostructures with tensile strain\cite{hung2023nonlinear}, or by creating nanotubes along the non-zero response direction\cite{mosse_tube}.
Furthermore, the possibility of inverting the SC direction by inverting the z-axis of Janus dichalcogenides and using their internal electric field to interact with closest layers 
endows these materials with extra tunability.

\section*{Acknowledgments}
C.A. and Y.M. acknowledge B. Demoulin and  A. Saul for the management of the computer cluster \emph{Rosa}. C.A. and Y.M. acknowledge ANR project COLIBRI No. ANR-22-CE30-0027. C.A. acknowledges funding from European Research Council MSCA-ITN TIMES under grant agreement 101118915. M.G. acknowledges funding from the UKRI Horizon Europe Guarantee funding scheme (EP/Y032659/1). C.A. acknowledge Davide Sangalli for helping developing the Yambo code.
\appendix
\section{Numerical evaluation of the current operator}
\label{app2}

In Eq.~\eqref{eq:current}, the current operator is written in term of a covariant $k$-derivative of the periodic part of the time-dependent valence bands. This derivative is numerically evaluated on a finite $k$-point grid as shown in Eq.~\eqref{eq:current2}. In this Appendix, we show how the formula for current is implemented in the  Yambo code~\cite{yambo}. All operators and wave-functions that appear in the equations of motions are expanded in the Kohn-Sham (KS) basis set $|u_{\kk,n}\rangle$. In this basis, the periodic part of the time-dependent valence bands  $| v_{\kk, n}\rangle$ and their covariant counterpart $|\widetilde{v}_{\kk i\sigma,n}\rangle$ reads: 
\bean
| v_{\kk, n}(t)\rangle &=& \sum_{j=1}^{\infty} c_{\kk,n,j}(t) | u_{\kk,j} \rangle\\
|\widetilde{v}_{\kk i\sigma,n}\rangle &=& \sum_{m=1}^{M} ( \tilde S^{-1}_{\kk i\sigma} )_{m,n} | v_{\kk i\sigma,m}\rangle 
\eean
where $M$ is the number of valence bands,  and $\tilde S_{\kk,i\sigma}$ is the overlap matrix between real-time valence states. The indices $m,n$ run on the valence bands.  $\{\kk i\sigma\} = \kk + \sigma \Delta\kk_i $ with $\sigma=\pm 1$ and $i$ labeling the displacement along the reciprocal lattice vectors in the $\kk$-grid.

The overlap matrix $S_{\kk,i\sigma}$ between time-dependent valence states reads: 
\bea
\[\tilde S_{\kk i\sigma} \]_{m,n} &=& \langle v_{\kk,m} (t) | v_{\kk i\sigma,n} (t)\rangle \\
\[\tilde S_{\kk i\sigma} \]_{m,n} &=&\sum_{j,l} c^*_{\kk,m,j}(t)  c_{\kk i\sigma,n,l}(t) (S_{\kk i\sigma})_{j,l} \\
\[S_{\kk i\sigma}\]_{j,l} &=&    \langle u_{\kk j} | u_{\kk i\sigma,l} \rangle  
\eea
We apply $\hat{H}_\kk^0$ to the bra and get:
\bea
\langle v_{\kk n}|\hat{H}_\kk^0&=&\sum_{j=1}^{\infty} c^*_{\kk,n,j}(t) \epsilon_{\kk,j} \langle u_{\kk,j} | \\
&=&\sum_{j=1}^{\infty} \bar c^*_{\kk,n,j}(t) \langle u_{\kk,j} | \nonumber 
\eea
where 
\be
\bar c^*_{\kk,n,j} = c^*_{\kk,n,j} \epsilon_{\kk,j} \label{cbar}. 
\ee
\\
Finally, we express also $|\widetilde{v}_{\kk i\sigma,n}\rangle$ in the KS basis set as:
\bean
|\widetilde{v}_{\kk i\sigma,n}\rangle &=& \sum_{m=1}^{M} ( \tilde S^{-1}_{\kk i\sigma} )_{m,n} | v_{\kk i\sigma,m}\rangle \\
&=& \sum_{m=1}^{M} ( \tilde S^{-1}_{\kk i\sigma} )_{m,n} \sum_{j=1}^{\infty} c_{\kk i\sigma,m,j}(t) | u_{\kk i\sigma,j} \rangle \\
&=& \sum_{m=1}^{M} ( \tilde S^{-1}_{\kk i\sigma} )_{m,n} \sum_{j,l=1}^{\infty} c_{\kk i\sigma,m,j}(t) \[S^*_{\kk,i\sigma}\]_{j,l} | u_{\kk,l} \rangle \\
&=&  \sum_{l=1}^{\infty} \tilde c_{\kk i\sigma,n,l}  | u_{\kk,l} \rangle
\eean
where
\be
\tilde c_{\kk i\sigma,n,l} (t) = \sum_{m=1}^{M} ( \tilde S^{-1}_{\kk,i\sigma} )_{m,n} \sum_{j=1}^{\infty} c_{\kk i\sigma,m,j}(t). \label{ctilde}
\ee
Using the definition of $\tilde c$ [Eq.~\eqref{ctilde}] and $\bar c$ [Eq.~\eqref{cbar}], we can write the bra-ket that appears in the current, Eq.~\ref{eq:current2}, as:
\bea
\langle v_{\kk n}|\hat{H}_\kk^0|\widetilde{v}_{\kk i\sigma,n}\rangle=\sum_{j=1}^{\infty} \bar c^*_{\kk,n,j}(t) \tilde c_{\kk i\sigma,n,j}(t) \label{eq:braket}
\eea
Using Eq.~\eqref{eq:braket} we can rewrite all Eq.~\eqref{eq:current2} in the KS basis.
\bibliographystyle{apsrev4-1}
\bibliography{nloptics.bib}
\end{document}